\documentclass[a4paper,12pt]{article}

\usepackage[dvips]{graphicx}
\usepackage{amsfonts}
\usepackage{epsfig,amssymb}

\begin{document}

\title{On the bifurcation and continuation of periodic orbits in the three-body problem}
\author{K. I. Antoniadou, G. Voyatzis, T. Kotoulas\\
Department of Physics, Aristotle University of Thessaloniki,\\ 54124 Thessaloniki, GR \\ kyant@auth.gr, voyatzis@auth.gr, tom\_1@otenet.gr}

\maketitle

\begin{flushleft}
Electronic version of an article published as [International Journal of Bifurcation and Chaos, Vol. 21, No. 8 (2011) $2211-2219$] [DOI: 10.1142/S0218127411029720] \copyright  [copyright World Scientific Publishing Company] [http://www.worldscinet.com/ijbc/]
\end{flushleft}

\begin{abstract}
We consider the planar three body problem of planetary type and we study the generation and continuation of periodic orbits and mainly of asymmetric periodic orbits.  Asymmetric orbits exist in the restricted circular three body problem only in particular resonances called ``asymmetric resonances''. However, numerical studies showed that in the general three body problem asymmetric orbits may exist not only for asymmetric resonances, but for other kinds, too.  In this work, we show the existence of asymmetric periodic orbits in the elliptic restricted problem. These orbits are continued and clarify the origin of many asymmetric periodic orbits in the general problem. Also, we illustrate how the families of periodic orbits of the restricted circular problem and those of the elliptic one join smoothly and form families in the general problem, verifying in this way the scenario described firstly by Bozis and Hadjidemetriou (1976). 
\end{abstract}

{\bf keywords} three body problem (TBP), periodic orbits, bifurcations.

\section{Introduction}
The three body problem (TBP), consisting of three bodies with masses $m_0$, $m_1$ and $m_2$ that interact with gravitational forces, has been widely studied in the literature. It is considered either as a simple dynamical system with complex behaviour or as a model for explaining the evolution of planets and other celestial bodies. In this work, we consider the planar case, where all bodies move on the same plane and the indices 0, 1 and 2  refer to the particular body. The simplest, yet not trivial, version of TBP is the {\em circular restricted TBP} (CRTBP), where we assume that one of the bodies is massless (say $m_2$=0) and the two massive bodies ($m_0\neq 0$, $m_1\neq 0$), called {\em primaries}, revolve in circular orbits on the plane. When the primaries revolve in elliptic orbits of eccentricity  $e_p$ (=$e_0=e_1)$ and period $T_p$ (=$T_0=T_1$), we have the {\em elliptic restricted TBP} (ERTBP). 

When the massless body is replaced by a massive one, we get the general planar model of the TBP (GTBP). Especially, when one body has very large mass in comparison with the other two bodies (say $m_0\gg m_1$, $m_0\gg m_2$), we get the so-called ``planetary'' GTBP. Continuation and existence of periodic orbits in this problem  has been studied many years ago by Hadjidemetriou (1975) and  Bozis and Hadjidemetriou (1976) and recently their results found a fruitful field of applicability in the dynamics of resonant extrasolar systems (e.g. Psychoyos and Hadjidemetriou, 2005; Haghighipour \emph{et al.}, 2003; Ferraz-Mello \emph{et al.}, 2005).

In most cases, periodic orbits are associated with resonant planetary motion and their determination in the general TBP can be based on the unperturbed circular problem (i.e. both planets are massless bodies). However, not all solutions can be constructed in this way. In Voyatzis \emph{et al.} (2009), it is shown that the main set of families of periodic orbits in the 2:1 resonance can be constructed by the continuation of periodic orbits from the restricted problem to the general one. In this scheme, the role of families of periodic orbits of the  restricted TBP is crucial.     

The origin of periodic orbits in the {\em circular restricted TBP} can be assigned to periodic orbits of the unperturbed problem. The structure of families of {\em symmetric} periodic orbits is well-known (see e.g. Bruno, 1994; H\'enon, 1997). Previous studies (e.g. Beaug\'e, 1994; Voyatzis  \emph{et al.}, 2005)) indicate that {\em asymmetric} periodic orbits exist only in resonances of the form $\frac{T_2}{T_1}=\frac{1}{q}$, $q\in \mathbb{N}$, called {\em asymmetric}, where for $q>1$ the massive body ($m_1\neq 0$) is the inner planet and the massless body ($m_2=0$) is the outer one. 

In the {\em elliptic restricted TBP}, families of periodic orbits bifurcating from periodic orbits of the circular problem also exist  having period $T=\frac{k}{\ell} T_p$, where $k$ and $\ell$ are prime integers (Broucke, 1969). Many periodic orbits for the elliptic model associated to the dynamics of asteroids and Kuiper belt objects have been computed  (e.g. Hadjidemetriou, 1999; Voyatzis  \emph{et al.}, 2005). All the periodic orbits found were symmetric. However, Voyatzis and Kotoulas (2005) found many bifurcation points along the families of periodic orbits and conjectured the generation of families of asymmetric orbits. In this paper, we compute and show the existence of such families in the elliptic restricted model and examine their continuation to the general model.

In the following section, we present our model and some basic notions on the description, existence and continuation of periodic orbits in the particular system. In section 3, we show how families of asymmetric periodic orbits can appear in the elliptic model and in section 4, how such families are continued to the general problem. Finally, in section 5, we conclude our results.

\section{The model of TBP and periodic orbits.}

\subsection{The model}
Considering the planetary TBP, consisting of a sun $S$ of mass $m_0$ and two planets $P_1$ and $P_2$ of masses $m_1\ll m_0$ and $m_2\ll m_0$, respectively, we define a rotating frame of reference in the following way (see Hadjidemetriou, 1975) : $O$ is the center of mass of $S$ and $P_1$, the axis $Ox$ is defined by the direction $S-P_1$ and the axis $Oy$ is vertical to $Ox$. Consequently, in the rotating frame, the position of the system is given by the coordinates $x_1$ (for $P_1$) and $x_2$, $y_2$ (for $P_2$), while the rotation of the frame is defined by the angular velocity $\dot\theta$ of the axis $Ox$ with respect to the inertial frame. The Lagrangian of the system can be written in the form
$$
{\cal L}=\frac{1}{2}(m_0+m_1) \left [\frac{m_1}{m_0}(1-\mu)^2(\dot
r^2+r^2\dot\theta^2)+\frac{m_2}{m} \Big( \dot x_2^2+\dot
y_2^2+\dot\theta^2 (x_2^2+y_2^2)
+2\dot\theta (x_2\dot y_2-\dot x_2 y_2) \Big)
\right ]+ V
$$
where $m=m_1+m_2+m_3$ is the total mass of the system,
$\mu=m_1/(m_0+m_1)$, $r=x_1/(1-\mu)$ is the distance of the Sun and $P_1$ and
$$
V=-\frac{G m_0 m_1}{r}-\frac{G m_1 m_2}{\sqrt{((1-\mu)r-x_2)^2+y_2^2}}-\frac{G m_0 m_2}{\sqrt{(\mu r+x_2)^2+y_2^2}}
$$
We remark that $\theta$ is ignorable and, therefore, the angular momentum $L=\partial {\cal L}/\partial\dot\theta$ is constant. Thus, the system can be reduced to three degrees of freedom. By taking the limit $m_2\rightarrow 0$ in the corresponding equations of motion, we obtain the equations of the elliptic restricted TBP. Furthermore, by considering the consistent solution $r=$const, $\dot\theta$=const, the equations reduce to the equations
of the circular restricted problem. In all cases, the system obeys the fundamental symmetry (H\'enon, 1997; Voyatzis and Hadjidemetriou, 2005)
\begin{equation} \label{EqSymmetry}
\Sigma: (t,x,y) \rightarrow (-t,x,-y).
\end{equation}

\subsection{Periodic orbits}
An orbit $\mathbf{X}(t)=(x_1(t),x_2(t),y_2(t),\dot x_1(t),\dot x_2(t), \dot y_2(t))$ of the general planar TBP is periodic of period $T$ if it satisfies the periodic conditions:
\begin{equation} \label{EqPeriodCond}
\begin{array}{ll}
\dot{x}_1(T)=\dot{x}_1(0)=0, & \\
x_1(T)=x_1(0), & \\ 
x_2(T)=x_2(0) ,& y_2(T)=y_2(0), \\
\dot{x}_2(T)=\dot{x}_2(0), & \dot{y}_2(T)=\dot{y}_2(0). 
\end{array}
\end{equation}

A periodic orbit is {\em symmetric} when it is invariant under the symmetry $\Sigma$ and {\em asymmetric} otherwise. An asymmetric periodic orbit $A$ is mapped by $\Sigma$ to another periodic orbit $A'$ which is the mirror image of $A$. Subsequently, all asymmetric periodic orbits appear in pairs $(A,A')$. 
  
The first of periodic conditions (\ref{EqPeriodCond}) and the conditon $\ddot x_1>0$ (or $\ddot x_1<0$)  define a Poincar\'e section in phase space. Studying the system on this section an asymmetric periodic orbit is assigned to a fixed point and is represented by a point in the 5-dimensional space of initial conditions
$$
\Pi_5=\{(x_1(0),x_2(0),y_2(0),\dot{x}_2(0),\dot{y}_2(0))\}.
$$     
A symmetric periodic orbit crosses perpendicularly the axis $Ox$ i.e., additionally to the condition $\dot x_1(0)=0$, we have $y_2(0)=0$ and $\dot x_2(0)=0$. Therefore, a symmetric periodic orbit can be represented by a point in the three dimensional space of initial conditions 
$$
\Pi_3=\{(x_1(0),x_2(0),\dot{y}_2(0))\}.
$$ 
The above periodicity conditions and the spaces $\Pi_5$ and $\Pi_3$ refer either to the GTBP or to the ERTBP. Considering the ERTBP, where the semimajor axis $a_1$ of $P_1$ is fixed by the normalization of units, the variable $x_1$ is given by the eccentricity of the planet $P_1$, namely $x_1(0)$=$a_1(1-e_1)$ or $x_1(0)$=$a_1(1+e_1)$ when $P_1$ is at pericenter or apocenter, respectively.  
    
In the CRTBP, $x_1$ is constant, defined by the normalization that is adopted for the system and $\dot x_1=0$. Thus, by considering now the section $y_2=0$, a symmetric orbit is defined as a point in the space of initial conditions $\Pi_2$=$\{$($x_2(0)$,$\dot{y}_2(0)$)$\}$ and the asymmetric one in the space $\Pi'_3$=$\{$($x_2(0)$, $\dot{x}_2(0)$, $\dot{y}_2(0)$)$\}$.  

\subsection{Continuation}
A periodic orbit of the CRTBP is continued mono-parametrically (e.g. by changing the energy integral) forming families (characteristic curves) in the spaces $\Pi_2$ or $\Pi'_3$ (symmetric or asymmetric, respectively). Along such a family  the period $T$ varies. In the non-autonomous ERTBP, the periodic orbits ought to have period which is an integer multiple of the period of primaries $T_p$, i.e. $T=k T_p$. A periodic orbit of the CRTBP can be continued to the ERTBP, if its period $T'$ is also equal to $T'=k T_p$ or $T'=k T_p/\ell$. In the latter case the generated periodic orbit  is assumed to be periodic with multiplicity $\ell$. 
     
If we consider a continuation to the space $\Pi_5$ with parameter the $x_1$ variable, which, as we have mentioned above, for the ERTBP is related to the eccentricity $e_1$ of the primaries, the periodicity conditions are written as 
\begin{equation} \label{EqCondPi5}
\mathbf{Y}(T;\mathbf{Y}(0))-\mathbf{Y}(0)=0,\quad \mathbf{Y}(t)=\{x_2(t),y_2(t),\dot x_2(t), \dot y_2(t)\}
\end{equation}
According to the implicit function theorem, the mono-parametric continuation to $\Pi_5$ is possible if the Jacobian of system (\ref{EqCondPi5}) has non-zero determinant, namely
\begin{equation} \label{EqContPi5} 
|\mathbf{J_4}(T)|=|\mathbf{\Delta_4}-\mathbf{I_4}|\neq 0, \quad  \mathbf{\Delta_4}=\frac{\partial \mathbf{Y}}{\partial \mathbf{Y}(0)},
\end{equation}
where $\mathbf{I_4}$ is the $4\times 4$ unit matrix. In case of symmetric periodic orbits, the periodicity conditions (\ref{EqCondPi5}) hold for $\mathbf{Y}(t)=\{x_2(t),\dot y_2(t)\}$ and Eq. (\ref{EqContPi5}) is rewritten by replacing index 4 by index 2.  In general, $\Delta_4$ has not a unit eigenvalue and, subsequently, (\ref{EqContPi5}) holds and continuation to $\Pi_5$ is generally possible. The cases where $|\mathbf{J_4}(T)|=0$ correspond to bifurcation points. 

In the ERTBP, the period $T$ in (\ref{EqContPi5}) is fixed. In the GTBP, $T$ is the time span between $k$ successive intersections with the Poincar{\'e} plane $\dot x_1=0$ and varies along the family of periodic orbits.  Continuation from the ERTBP to GTBP is possible when we vary the mass $m_2$ and keep fixed the variable $x_1$. Also, in this case, we obtain again the periodicity and continuation conditions (\ref{EqCondPi5}) and (\ref{EqContPi5}), respectively. 

In the GTBP, the initial conditions of a periodic orbit correspond to particular eccentricity values $e_1$ and $e_2$, called {\em osculating} eccentricities, for the planetary orbits.  For the representation of families of periodic orbits in the space $\Pi_5$ it is convenient to use the plane $e_1 - e_2$. In this plane, a family of asymmetric orbits coincides with its relative family consisting of {\em mirror image} orbits. Also, the families of the CRTBP lie on the axis $e_1$ or $e_2$, since in this model one of the planet, $P_2$ or $P_1$ respectively, has zero eccentricity.

\section{Asymmetric periodic orbits in the ERTBP}
The generation and the numerical continuation of symmetric periodic orbits in the ERTB have been studied in applications concerning the resonant evolution of minor celestial bodies (Hadjidemetriou 1999; Voyatzis and Kotoulas, 2005).  As far as we know, any asymmetric orbits have been given in literature for the ERTBP. We propose that such asymmetric periodic orbits can be obtained after two different types of bifurcation discussed in the following.

\subsection{Type I bifurcation.} Along the symmetric families of the ERTBP the linear stability may change. At these cases we have critically stable periodic orbits where $\det|\mathbf{J_4}|=0$ and, generally,  $\det|\mathbf{J_2}|\neq 0$. Therefore, those orbits should be bifurcation points in $\Pi_5$ for asymmetric families of periodic orbits. We note that the change in linear stability is a quite common fact in the ERTBP (Voyatzis and Kotoulas, 2005) and, therefore, the families of asymmetric periodic orbits should play an important role in its dynamics.

\begin{figure}
\begin{center}
\includegraphics[width=10cm]{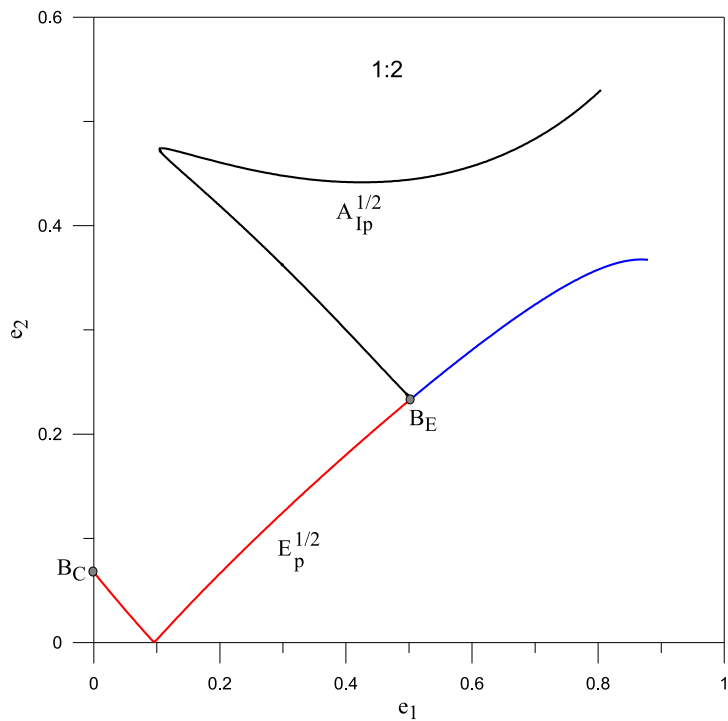}
\end{center}
\caption{The 1:2 resonant symmetric family $E^{1/2}_p$ of the ERTBP and the biffurcation of the asymmetric family $A^{1/2}_{Ip}$.}
\label{Fig21BtypeI}
\end{figure}

In Fig. \ref{Fig21BtypeI}, a 1:2 family of symmetric periodic orbits $E^{1/2}_p$ is presented which starts from the bifurcation point $B_C$, as an unstable family of period $2T_p$. Point $B_C$ belongs to an unstable segment of a symmetric family of the CRTBP and family $E^{1/2}_p$ starts also as unstable and continues parametrically by increasing $e_1$. At $e_1=0.5022$ the family changes stability type and becomes stable. At this point, $B_E$, we have a bifurcation of the asymmetric family $A^{1/2}_{Ip}$, which starts as stable.  The family $E^{1/2}_a$, which corresponds to the apocenter initial position of the planet $P_1$, is unstable for all values of $e_1$ and, therefore, it does not show a bifurcation of this type. 

\begin{figure}
\begin{center}
\includegraphics[width=10cm]{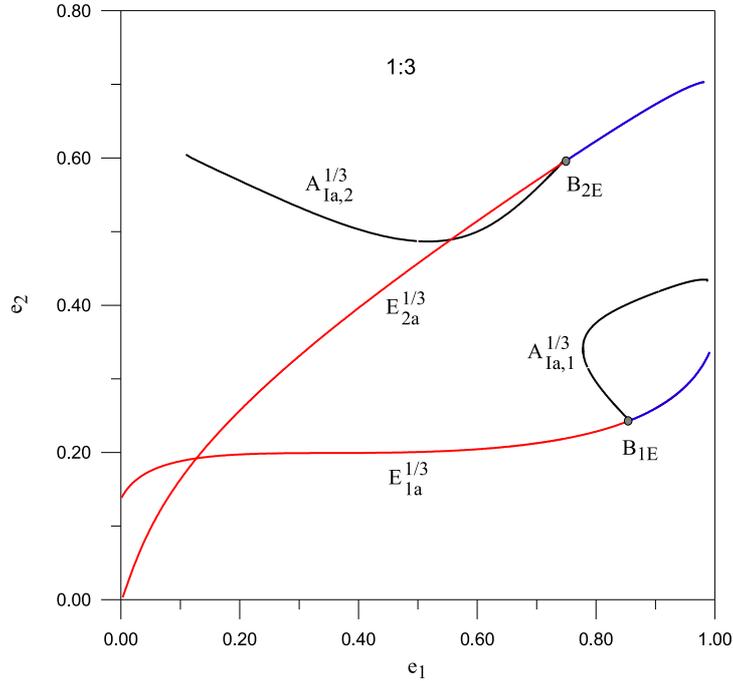}
\end{center}
\caption{The 1:3 resonant symmetric families $E^{1/3}_{1a}$ and $E^{1/3}_{2a}$ of the ERTBP and the biffurcation of the asymmetric families $A^{1/3}_{Ia,1}$ and $A^{1/3}_{Ia,2}$.}
\label{Fig31BtypeI}
\end{figure}

A similar structure as above is met in the 1:3 resonance (see Fig. \ref{Fig31BtypeI}). In this case, we have found two symmetric families of ERTBP that show bifurcations of type I. The first family $E^{1/3}_{1a}$,  bifurcates from the 1:3 resonant symmetric family of the CRTBP, while the second one  $E^{1/3}_{2a}$ starts from the circular family of the CRTBP. Both of them start as unstable, but they change stability at points $B_{1E}$ and $B_{2E}$, respectively. At these points, we have the bifurcation of the asymmetric families $A^{1/3}_{Ia,1}$ and $A^{1/3}_{Ia,2}$. 

\begin{figure}
\begin{center}
\includegraphics[width=10cm]{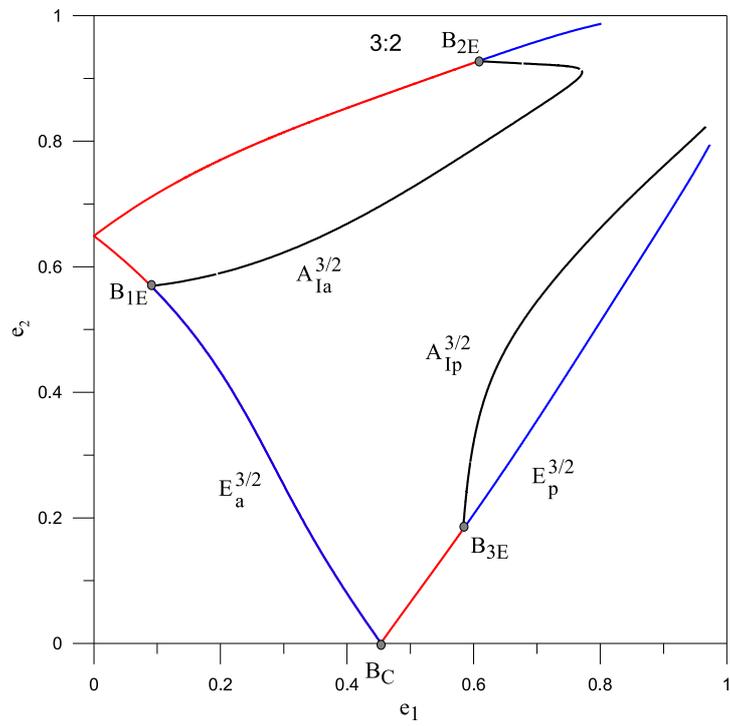}
\end{center}
\caption{The 3:2 resonant symmetric families $E^{3/2}_p$ and $E^{3/2}_a$ of the ERTBP and the biffurcation of the asymmetric families $A^{3/2}_{Ip}$ and $A^{3/2}_{Ia}$.}
\label{Fig32BtypeI}
\end{figure}

Bifurcations of type I are also found for the 3:2 resonance. This resonance is neither  ``asymmetric'' nor ``external'', i.e. the massless planet $P_2$ revolves inside the orbit of the massive planet. In these cases, the CRTBP has no asymmetric families. Particularly, the 3:2 resonance has a stable family of symmetric periodic orbits which shows a bifurcation point to the ERTBP at $e_1=0.4529$ ($e_2=0$). Two families of symmetric periodic orbits, $E^{3/2}_p$ and $E^{3/2}_a$, originate from this point and both of them start as stable. This is shown in Fig. \ref{Fig32BtypeI}. Both families change their stability and the bifurcation points, $B_{1E}$, and $B_{3E}$, for asymmetric families are obtained. These families are indicated as $A^{3/2}_{Ip}$ and $A^{3/2}_{Ia}$ in Fig. \ref{Fig32BtypeI}. We note that the family $E^{3/2}_a$ turns again to become stable showing a second bifurcation point $B_{2E}$. This point does not generate a new family, but it may be assumed as the termination point of family $A^{3/2}_{Ia}$.
     
\subsection{Type II bifurcation.} Considering the asymmetric families of periodic orbits in the CRTBP, which exist for resonances in the form of $T_2/T_1=1/q$, where the primary planet is the inner one (Beaug\'e, 1994, Voyatzis \emph{et al.}, 2005), we compute the period $T$ of orbits. As we have mentioned the orbits of these families are represented in the space $\Pi'_3$=$\{$($x_2(0)$, $\dot{x}_2(0)$, $\dot{y}_2(0)$)$\}$. If $T=k T_p$ then they consist bifurcation points and periodic orbits are generated in the ERTBP by continuation after varying $x_1$ or, equivalently, $e_1$. Since at the bifurcation $\dot x_2\neq 0$, in general, the orbits after continuation should have also $\dot x_2\neq 0$ and, subsequently, they are asymmetric. 

\begin{figure}
\begin{center}
$\begin{array}{ccc}
\includegraphics[width=7cm]{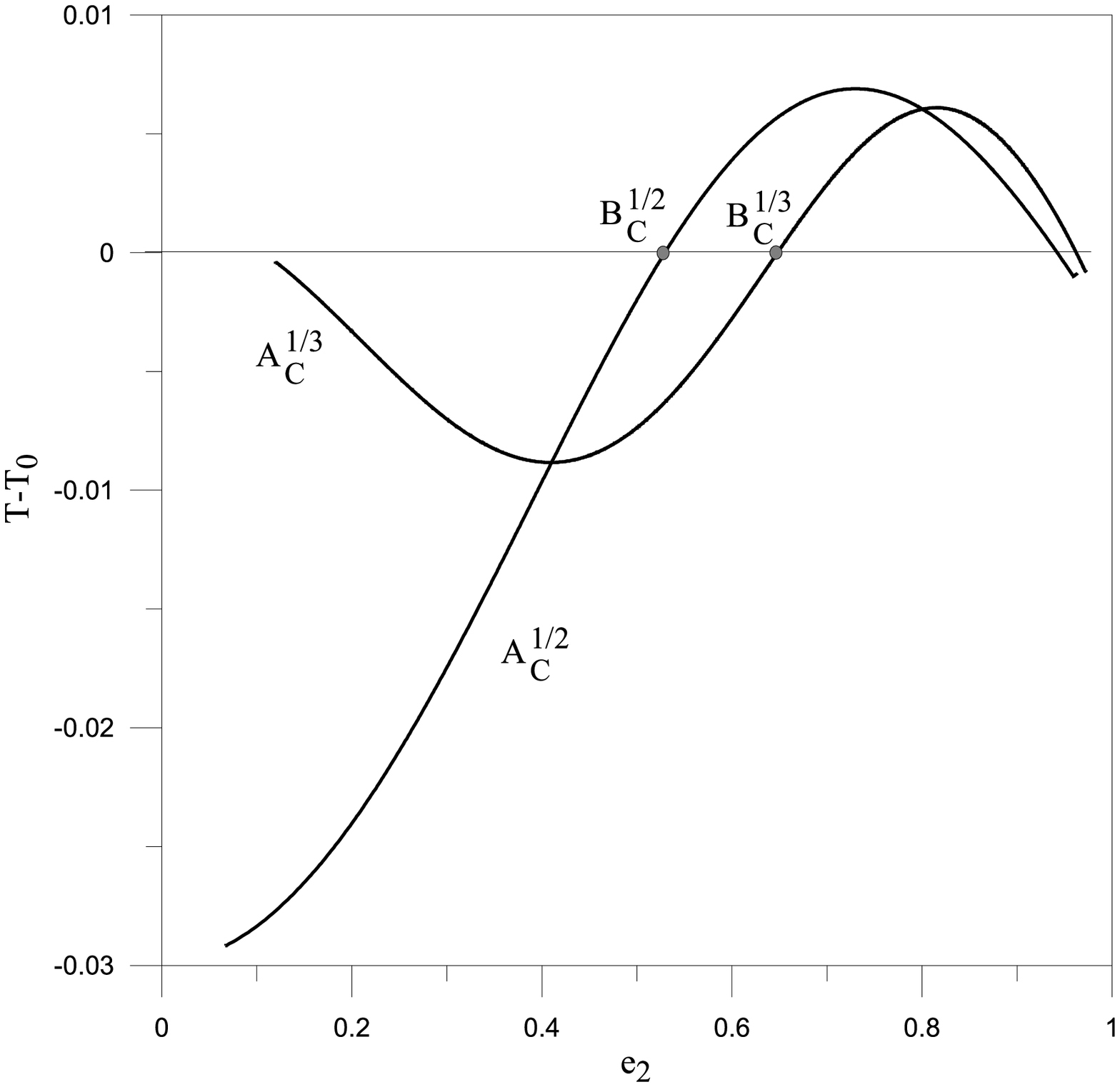} & \qquad \qquad&
\includegraphics[width=7cm]{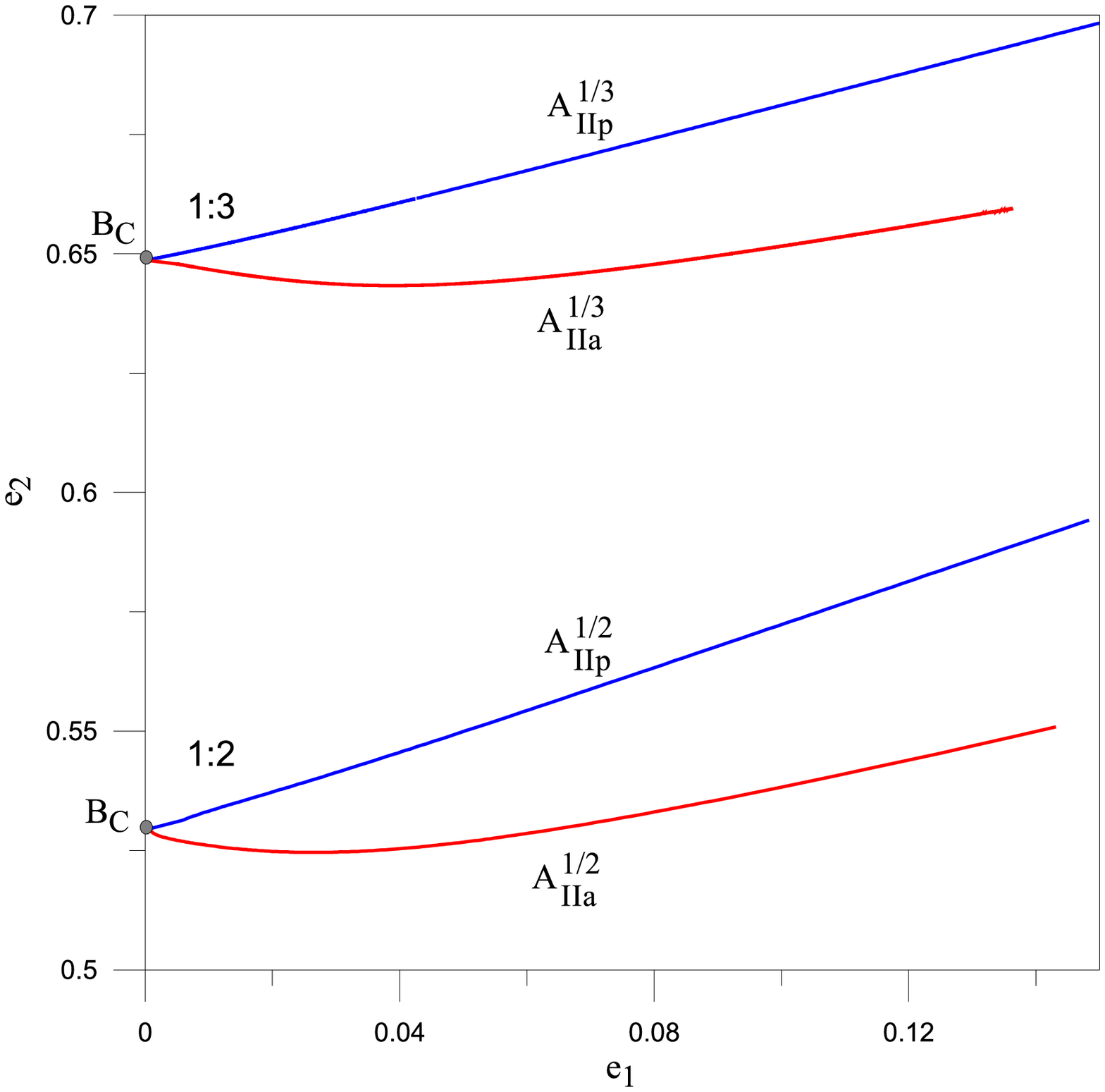}\\
\textnormal{(a)} & \qquad\qquad & \textnormal{(b)} 
\end{array} $
\end{center}
\caption{(a) The variation of the difference $T-T_0$ along the asymmetric families of the CRTBP. The biffurcation points correspond at $T-T_0=0$. 
(b) Type II bifurcation and continuation of 1:2 and1:3 resonant asymmetric families in the ERTBP.}
\label{FigBifurII}
\end{figure}

In Fig. \ref{FigBifurII}a,  we show the variation of the period along the 1:2 and 1:3 resonant families of asymmetric periodic orbits in CRTBP. In particular, we present the difference $T-T_0$, where $T$ is the period of the orbits along the family and $T_0=2T_p$ or $T_0=3T_p$ for the 1:2 and 1:3 resonant families, respectively. Bifurcations for families in the ERTBP correspond to the periodic orbits where $T-T_0=0$. We can observe the existence of two bifurcation points for each family. We indicate the two of them and present the asymmetric families which are generated from them in Fig. \ref{FigBifurII}b.  From each bifurcation point two asymmetric families originate, according to whether the planet $P_1$ is initially at pericenter or apocenter. Thus, for the resonance 1:2, we get the asymmetric families $A^{1/2}_{IIp}$, which is linearly stable, and $A^{1/2}_{IIa}$, which is linearly unstable. Similarly we obtain the 1:3 resonant families $A^{1/3}_{IIp}$ and $A^{1/3}_{IIa}$ which are stable and unstable, respectively.

\section{Continuation from the restricted to the general problem}
Considering a periodic orbit of the restricted problem, circular or elliptic, we can perform continuation by giving and increasing the mass of the initially massless body. The continuation of symmetric periodic orbits from the restricted to the general problem has been studied by Bozis and
Hadjidemetriou (1976). In the paper of Voyatzis \emph{et al.} (2009) relative computations were performed for the 1:2 resonance. In this
study, we have extended our computations for the 1:3 and 3:2 resonances and present two schemes of continuation.
           
\subsection{Continuation Scheme I.}
We mentioned in section 2.2 that the periodic conditions of the ERTBP and the GTBP are equivalent. Also, the determinants $|J(T)|$ of the periodic conditions, for either the symmetric or asymmetric orbits, are continuous functions with respect to the planetary masses. Consequently, in the general case, where $|J(T)\neq 0|$, all families of periodic orbits of the ERTBP are continued in the general problem by varying the masses. 

\begin{figure}
\begin{center}
$\begin{array}{ccc}
\includegraphics[width=7cm]{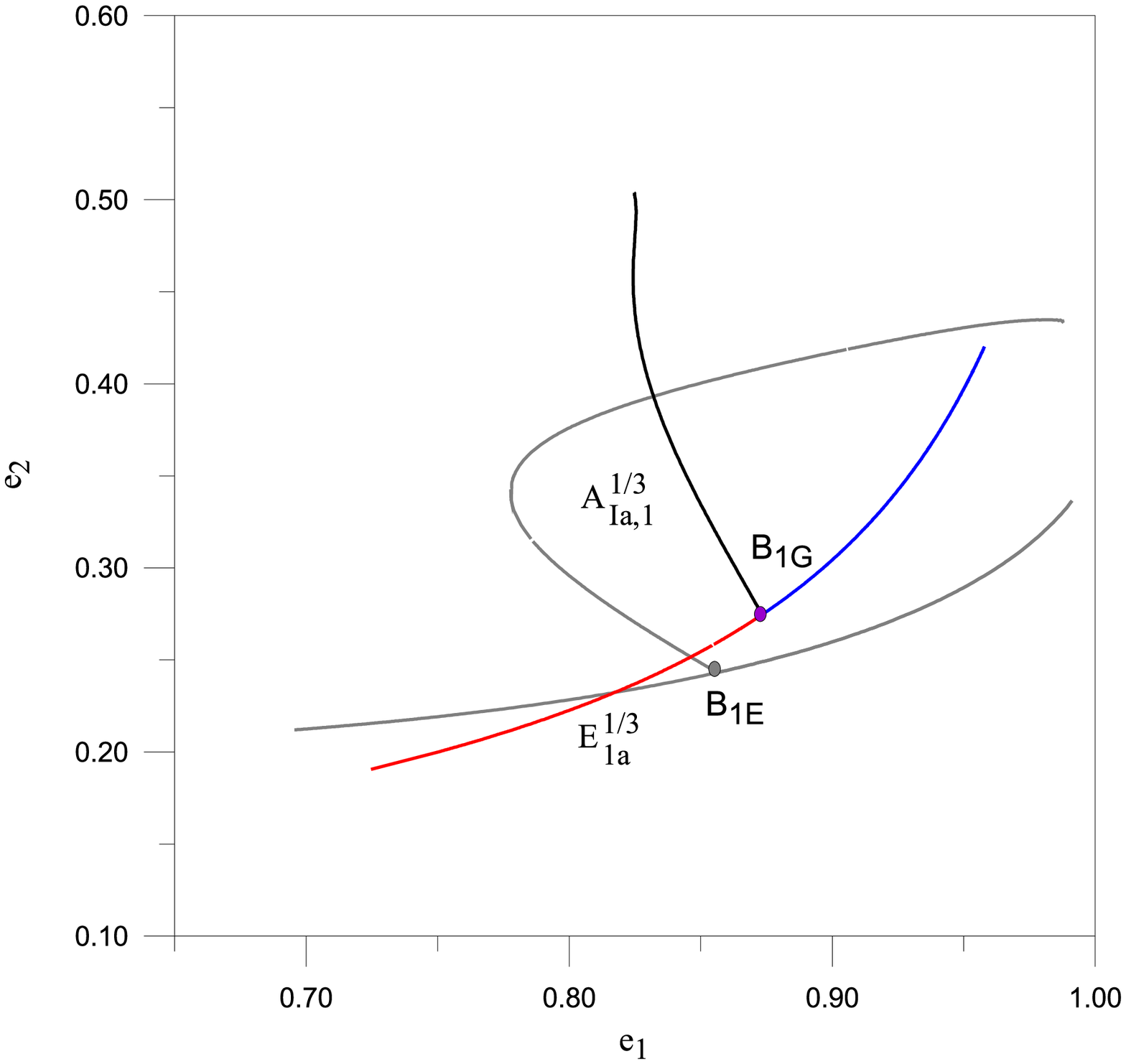} & \qquad \qquad&
\includegraphics[width=7cm]{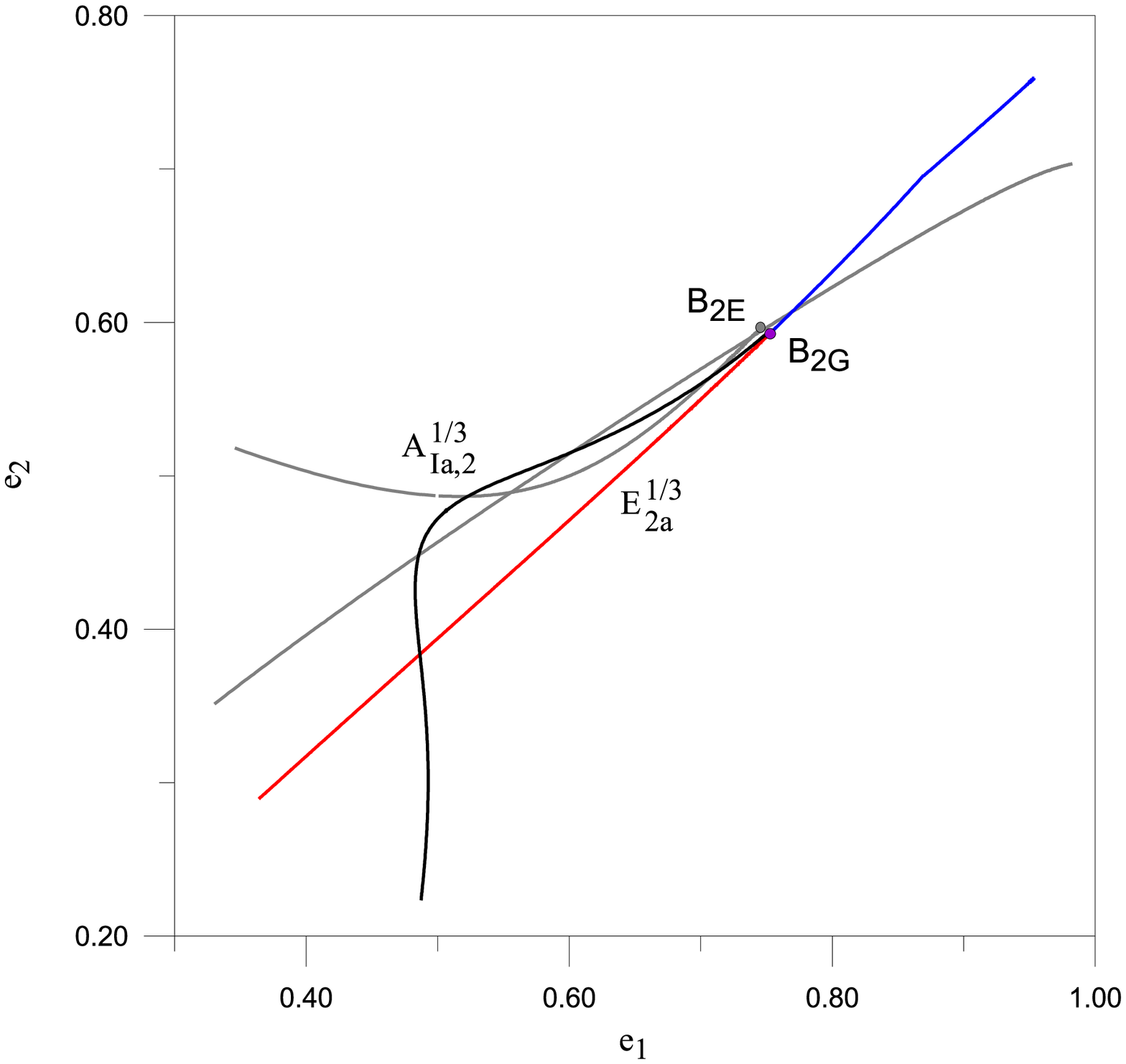}\\
\textnormal{(a)} & \qquad\qquad & \textnormal{(b)} 
\end{array} $
\end{center}
\caption{(a) The families $E^{1/3}_{1a}$ and $A^{1/3}_{Ia,1}$ (gray curves) of the ERTBP ($m_2=0$) and their continuation to the GRTBP at $m_2=0.003$. (b) The same for the families $E^{1/3}_{2a}$ and $A^{1/3}_{Ia,2}$}
\label{Fig13ContI}
\end{figure}

Particularly, the families evolved in bifurcation of type I  do not change their topological structure after giving non zero mass to the small body. In Figs. \ref{Fig13ContI}a and \ref{Fig13ContI}b, we show the shape modification of the symmetric families $E^{1/3}_{1a}$ and $E^{1/3}_{2a}$ and the asymmetric ones $A^{1/3}_{Ia,1}$ and $A^{1/3}_{Ia,2}$, when they are continued to the general problem. We present them in the plane of eccentricities for $m_2=0$ and $m_2=0.003$.   

\subsection{Continuation Scheme II} 
According to Bozis and Hadjidemetriou (1976) all periodic orbits of the CRTPB are continued in the general problem except those having period $T=k T_p$, where, $k=1,2,...$ and $T_p$ the period of the circular orbit of primaries. These orbits are exactly the bifurcation orbits of the families in the ERTBP. Therefore, at this point, the topology of the families is not the same with that described in the scheme I, but a gap is formed, as we show in the following.

\begin{figure}
\begin{center}
\includegraphics[width=10cm]{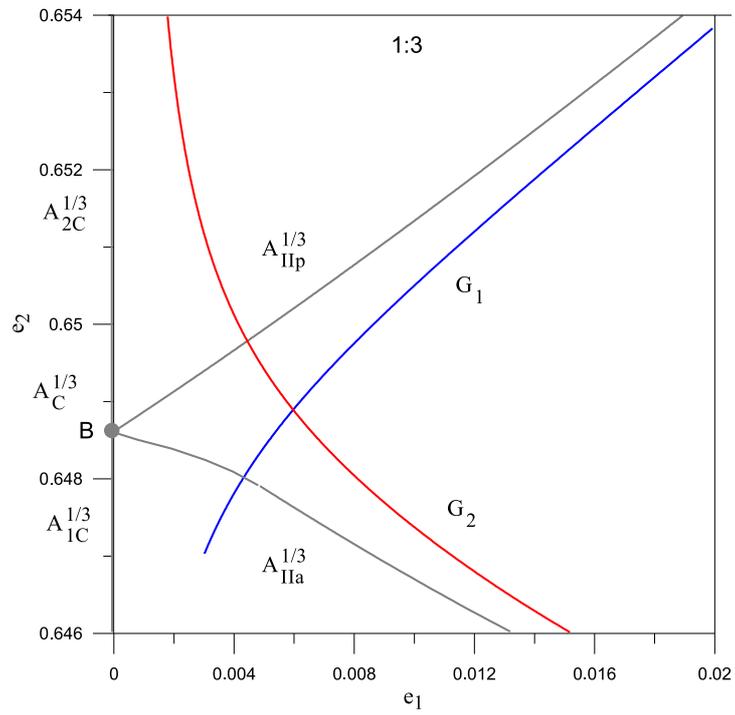}
\end{center}
\caption{The generation of families of asymmetric orbits in the GTBP ($G_1$ and $G_2$) after the combination of the family $A_{C}^{1/3}$ of the CRTBP with the families $A_{IIp}^{1/3}$ and $A_{IIa}^{1/3}$ of the ERTBP.}
\label{Fig13ContII}
\end{figure}

We present the continuation of periodic orbits at $T=k T_p$ by an example shown in Fig. \ref{Fig13ContII}. We consider the asymmetric families of the ERTBP $A^{1/3}_{Ip}$ and $A^{1/3}_{Ia}$, which bifurcate from the asymmetric family $A^{1/3}_c$ and are shown in Fig. \ref{FigBifurII}b for $m_2=0$. We denote by $B$ the periodic orbit in $A^{1/3}_c$ which has period $T=3 T_p$ and gives the bifurcation point for the families $A^{1/3}_{Ip}$ and $A^{1/3}_{Ia}$. The point $B$ divides the family  $A^{1/3}_c$ in two segments denoted by $A^{1/3}_{1c}$ and $A^{1/3}_{2c}$. When $m_2$ takes a non zero value the family $A^{1/3}_c$ breaks at point $B$. The two segments $A^{1/3}_{1c}$ and $A^{1/3}_{2c}$ separate from each other and join smoothly the families $A^{1/3}_{IIp}$ and $A^{1/3}_{IIa}$, respectively. So, we get two asymmetric families for $m_2\neq 0$, denoted in Fig. \ref{Fig13ContII} by $G_1$ and $G_2$. At the point $B$ a gap is formed.        

\begin{figure}
\begin{center}
\includegraphics[width=10cm]{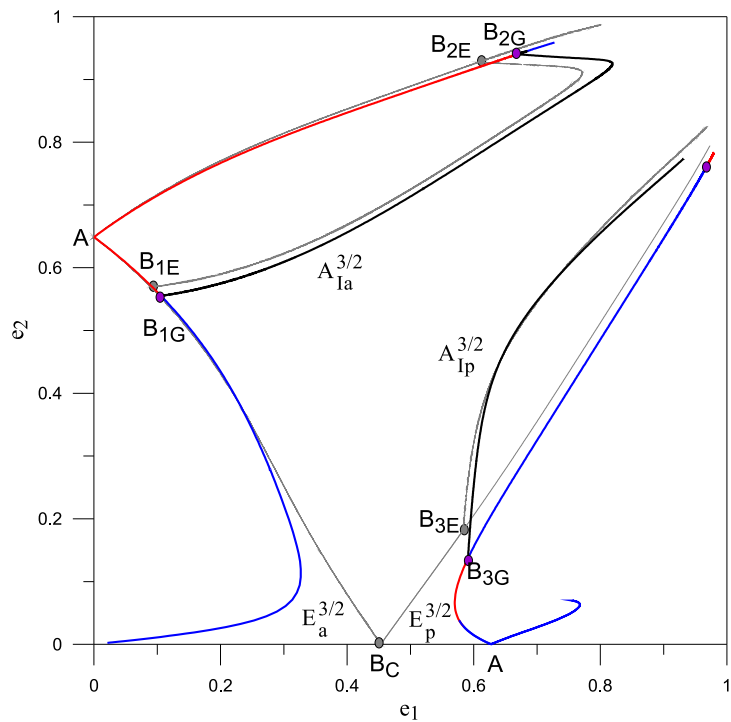}
\end{center}
\caption{The generation of families of the GTBP in the 3:2 resonance. Gray thin curves correspond to the families of the restricted problem ($m_1=0$) and the wider ones correspond to $m_1=0.0001$. Asymmetric families follow the continuation scheme I while the symmetric ones follow the continuation scheme II.}
\label{Fig32Cont}
\end{figure}

\subsection{Continuation of 3:2 resonant orbits}
The above continuation schemes are present in the internal resonance 3:2 as it is shown in Fig. \ref{Fig32Cont}. We remind that in this case the primary planet is $P_2$, i.e. $m_2\neq 0$ and $m_1=0$ in the restricted problem.  At the bifurcation points $B_{1E}$, $B_{2E}$ and $B_{3E}$,  we have the continuation scheme I by increasing the mass $m_1$. The position and shape of the families of the ERTBP and GRTBP differ slightly as long as $m_1$ is small. These families explain the origin of the 3/2 resonant asymmetric families found by Michtchenko \emph{et al.}, (2006) using an averaged model of the planetary three body problem. In the 3/2 resonance, we do not have asymmetric orbits generated from bifurcation points of type II. However, the symmetric family of the circular problem, lying on the axis $e_2=0$ in Fig. \ref{Fig32Cont}, has an orbit of period $T=2 T_p$ (point $B_C$). Subsequently, we get the symmetric families denoted by $E^{3/2}_p$ and $E^{3/2}_a$ which are continued in the general problem ($m_2\neq 0$) according to the scheme II.  

We remark that, when the masses increase enough, the families may intersect and new bifurcations arise in the framework of the general problem changing drastically the topological structure of families. These bifurcations are called ``collision bifurcations'' and are studied in Voyatzis \emph{et al.} (2009).

\section{Conclusions}

We have studied some new types of bifurcation of periodic orbits in the planar three body problem  consisting of a heavy body (star) and two bodies of small or negligible mass (planets). When one planet has zero mass (massless body) but the second one is massive (primary body), we have the simplest models of the circular (CRTBP) or the elliptic (ERTBP) restricted three body problem.  Considering these models in an appropriate rotating frame, we can compute resonant periodic orbits and then perform continuation to the general problem where both planets are massive.
 
It is known that periodic orbits in the ERTBP are generated from periodic orbits of the CRTBP and are continued parametrically, by varying the eccentricity of the primary body, $e_p$.  These orbits are generally symmetric. In this paper, we have shown the existence of families of asymmetric periodic orbits in the ERTBP.  There are two types of bifurcations of such orbits, which we called as type I and II bifurcations.  In a type II bifurcation, the asymmetric orbits of the ERTBP are generated from an asymmetric periodic orbit of the CRTBP. It is known, that such periodic orbits in the CRTBP exist only in the resonances of the form $1/q$ and therefore such families of periodic orbits exist for the RTBP only in these resonances. However, in a bifurcation of type I, the asymmetric families are generated from symmetric periodic orbits of the ERTBP. Actually, these bifurcation points are critical orbits with respect to their linear stability and can be met in any resonance. We presented such families for the 1:2 and 1:3 resonance and  also for the resonance 3:2, which is not of the form $1/q$. 

The families of the CRTBP or ERTBP are continued by varying the planetary masses. So, they become families of the general planar three body problem. 
Their continuation with respect to the planetary masses can follow two different schemes. The continuation scheme I holds for the families related to  type I bifurcation in the ERTBP. Namely, these families continue smoothly by varying the mass preserving their topological structure in phase space for adequately small planetary masses.  In the continuation scheme II, families of both the CRTBP and ERTBP models involve in order to generate a family in the GTBP.    

Following the generation and the continuation of the above mentioned families we complete the net of families of periodic orbits in the planar general three body model and we explain their origin. Periodic orbits of the TBP correspond to the so-called "exact resonances"  in the planetary dynamics and are important since an exosolar planetary system can be trapped in such orbits after a physical dissipation process.

\end{document}